\documentclass[aps,showpacs,prl,twocolumn,letterpaper]{revtex4}
\usepackage{amsmath,amssymb}

\begin{document}
\title{On the optimal contact potential of proteins}
\author{Akira R. Kinjo}
\affiliation{Institute for Protein Research, Osaka University, Suita, Osaka, 565-0871, Japan}
\email{akinjo@protein.osaka-u.ac.jp}
\author{Sanzo Miyazawa}
\affiliation{Faculty of Technology, Gunma University, Kiryu, Gunma 376-8515, Japan.}

\date{\today}
\begin{abstract}
We analytically derive the lower bound of the total conformational energy
of a protein structure by assuming that the total conformational energy is 
well approximated by the sum of sequence-dependent pairwise contact energies. 
The condition for the native structure achieving the 
lower bound leads to the contact energy matrix that is 
a scalar multiple of the native contact matrix, i.e., the 
so-called G\=o potential. We also derive spectral relations 
between contact matrix and energy matrix, and approximations
related to one-dimensional protein structures.
Implications for protein structure prediction are discussed.

\end{abstract}
\maketitle
\section{Introduction}
Proteins' biological functions are made possible by their precise 
three-dimensional (3D) structures, and each 3D structure is determined by 
its amino acid sequence through the laws of thermodynamics~\cite{Anfinsen1973}.
Therefore, predicting protein structures from their amino acid sequences
is important not only for inferring proteins' 
biological functions, but also for understanding how 3D structures 
are encoded in such one-dimensional information as amino acid sequence.
The problem of protein structure prediction is naturally cast as an 
optimization problem where a potential function is minimized. Given an 
appropriate potential function, conformational optimization should yield 
the native structure as the unique global minimum conformation of the 
potential function.
Thus, the problem has been traditionally divided into two sub-problems:
One is to establish an appropriate potential function~\cite{HaoANDScheraga1999}, and the other is to 
develop the methods to efficiently search the vast conformational space of
a protein~\cite{MitsutakeETAL2001}. 
Among various forms of effective energy functions, statistical contact 
potentials~\cite{MiyazawaANDJernigan1985,MiyazawaANDJernigan1996} have been 
widely used.
In this Letter, we exclusively treat a class of such contact potentials, 
neglecting other contributions such as electrostatics and local interactions.
Accordingly, a protein conformation is represented as a contact matrix in which 
the $(i,j)$ element is 1 if the residues $i$ and $j$ are in contact in space,
otherwise it is 0.
Although the contact matrix is a coarse-grained representation of protein 
conformation, it has been known that the contact matrix contains sufficient 
information to recover the three-dimensional (native) structure of 
proteins~\cite{VendruscoloETAL1997}. 
It is noted that, for the lattice model of 
proteins~\cite{TaketomiETAL1975}, these representations 
of protein conformation and energy function are exact.

\section{Theory}
\subsection{Lower bound of contact energy}
Our fundamental assumption is that the conformational energy of a protein
can be somehow expressed in terms of a contact matrix.
Now let us assume that the total energy of a protein can be well approximated 
by the sum of pairwise contact energies between amino acid residues, and that
each pairwise contact energy can be decomposed into a sequence-dependent 
term and a conformation-dependent term.
The sequence-dependent term is expressed as a matrix 
$\mathcal{E}(S) = (\mathcal{E}_{ij})$ which we call the contact energy matrix, 
or $E$-matrix for short. Each element $\mathcal{E}_{ij}$ of the $E$-matrix 
represents the energy between the residues $i$ and $j$ when they are in contact.
This form of the $E$-matrix is a very general one: Each
element, $\mathcal{E}_{ij}$, may depend on the entire sequence, $S$, or 
it may depend only on the types of the interacting amino acid residues, 
$i$ and $j$, as in the conventional contact potentials.
The conformation-dependent term is expressed as another 
matrix $\Delta(C) = (\Delta_{ij})$ which we call the contact matrix, or 
$C$-matrix. Each element $\Delta_{ij}$ of the $C$-matrix assumes a value of 
either 1 or 0, depending on the residues $i$ and $j$ are in contact or not,
respectively.
Hence the total energy $E(C,S)$ of a protein
of sequence $S$ of $N$ residues and having conformation $C$ is given by
\begin{eqnarray}
  \label{eq:ene}
  E(C,S) &=& \frac{1}{2}\sum_{i=1}^{N}\sum_{j=1}^{N} \mathcal{E}_{ij}(S)\Delta_{ij}(C) \\
\label{eq:inner}
&=& \frac{1}{2}[\mathcal{E}(S), \Delta{}(C)]
\end{eqnarray}
where $[\cdot , \cdot]$ denotes the Frobenius inner product between 
two matrices~\cite{MatrixAnalysis,TopicsInMatrixAnalysis}.
Based on this assumption, we derive the lower bound for the conformational 
energy and the conditions for the native structure and $E$-matrix to achieve
the bound.

The Frobenius inner product leads to the matrix $l_2$ norm defined as, for 
a matrix $M$, $\|M\| \equiv [M, M]^{1/2} = (\sum_{i,j}M_{ij}^2)^{1/2}$. 
In the case of $C$-matrix, since $\Delta_{ij} = 0$ or $1$, we have 
\begin{equation}
  \label{eq:tot-cn}
\|\Delta(C)\|^2 = 2N_c(C)
\end{equation}
where $N_c \equiv (1/2)\sum_{i,j}\Delta_{ij}$ is the total number of contacts. 
As for any inner products, the Frobenius inner product satisfies the 
Cauchy-Schwarz inequality ($|[A,B]| \leq \|A\|\|B\|$) from which we have
\begin{equation}
  \label{eq:cs-ineq}
  [\mathcal{E}, \Delta] \geq -\|\mathcal{E}\|\|\Delta\|
\end{equation}
where the equality holds if and only if 
\begin{equation}
  \label{eq:go-pot}
  \mathcal{E} = \varepsilon \Delta
\end{equation}
for some scalar $\varepsilon < 0$. 
Although the inequality (Eq. \ref{eq:cs-ineq}) holds for any pair of matrices,
we now regard it as the lower bound for conformational energy for a 
given $E$-matrix. 
For simplicity, we first consider the energy minimization problem 
for conformations with $\|\Delta(C)\|$ fixed to the value of the native 
conformation. 
It is desirable for the native conformation to satisfy the lower bound 
and hence its condition Eq. (\ref{eq:go-pot}).
If the native conformation indeed satisfies the condition 
Eq. (\ref{eq:go-pot}), then the elements of the $E$-matrix is either 0 
or $\varepsilon$ so that only the contacts present in the native conformation 
are stabilizing. Thus, the native conformation satisfying Eq. (\ref{eq:go-pot}) is actually a GMEC among any
conformations with arbitrary values of $\|\Delta(C)\|$.
An $E$-matrix that satisfies Eq. (\ref{eq:go-pot}) for the native 
$C$-matrix is a kind of the so-called G\=o potential~\cite{Go1983,Takada1999}
which has been essential for studying the protein folding problem.
At this point, it is still possible that the native structure is not the 
unique GMEC. For example, if a conformation contains all the native contacts 
together with some other contacts, this conformation has the same energy as 
the native conformation.
In order for a native conformation to be the unique GMEC, it is required that
the total number of contacts of the native conformation is larger than 
that of any other conformations that contain all the native contacts. 
From the relation Eq. (\ref{eq:tot-cn}), maximizing the total number of 
contacts is equivalent to maximizing the norm of the $C$-matrix, 
which in turn implies the minimization of the right-hand side of 
Eq. (\ref{eq:cs-ineq}). 
To summarize, for a given $E$-matrix, $\mathcal{E}(S)$, of a protein,
its native conformation, $C_{n}$, achieves the lower bound in 
Eq. (\ref{eq:cs-ineq}) if and only if 
$\mathcal{E}(S) = \varepsilon\Delta(C_n)$ for some $\varepsilon < 0$,
and such native structure is the unique GMEC if and only if
$\|\Delta(C_n)\|$ is the maximum of all possible conformations that contain
all the native contacts. Note that the former condition is a relation between 
$E$-matrix and $C$-matrix whereas the latter is a condition for a native 
structure to satisfy.
The magnitude of $\varepsilon$ is not specified here, 
but it should be determined by other factors such as the folding temperature.
It should be noted that a native structure can be the unique GMEC without 
achieving the lower bound of Eq. (\ref{eq:cs-ineq}).
Such a case is made possible either by the limitation of the 
conformational space imposed by other steric factors such as chain 
connectivity or excluded volumes, or by inherent inconsistencies of 
the $E$-matrix so that no plausible conformations are allowed to satisfy 
the lower bounds. 

\subsection{Spectral relations}
To examine more closely how the lower bound can be achieved, we next derive 
a more generous lower bound in a more restricted case. 
First, the $C$-matrix is decomposed as 
\begin{equation}
  \label{eq:c-decomp}
  \Delta = \sum_{\alpha=1}^{N}\sigma_{\alpha}\mathbf{u}_{\alpha}\mathbf{v}_{\alpha}^{T}
\end{equation}
where $\sigma_{\alpha}$ is the $\alpha$-th singular value and 
$\mathbf{u}_{\alpha}$ and $\mathbf{v}_{\alpha}$ are the corresponding left 
and right singular vectors, respectively.
$U = (\mathbf{u}_1, \cdots , \mathbf{u}_N)$ 
and $V = (\mathbf{v}_1, \cdots , \mathbf{v}_N)$ are
orthogonal matrices. The singular components are sorted in decreasing 
order of the singular values: $\sigma_{1} \geq \cdots \geq \sigma_{N} 
(\geq 0)$. 
Since $\Delta$ is real symmetric, the singular values are the absolute values 
of the eigenvalues of $\Delta$, and the singular vectors are such that 
$\mathbf{u}_{\alpha} = \pm \mathbf{v}_{\alpha}$ where the sign corresponds to 
that of the respective eigenvalue.
Next, the $E$-matrix is decomposed in the same manner as
\begin{equation}
  \label{eq:e-decomp}
  \mathcal{E} = \sum_{\alpha=1}^{N}\tau_{\alpha}\mathbf{x}_{\alpha}\mathbf{y}_{\alpha}^{T}
\end{equation}
where $\tau_{\alpha}$ are singular values, and $\mathbf{x}_{\alpha}$ and 
$\mathbf{y}_{\alpha}$ are left and right singular vectors, respectively.
Since $\mathcal{E}$ is also real symmetric, the singular components have 
the same properties as the $C$-matrix $\Delta$.
Noting that $[\mathcal{E},\Delta] = \mathrm{tr} (\mathcal{E}\Delta^{T})$, 
von Neumann's trace theorem~\cite{TopicsInMatrixAnalysis} leads to
the following inequality:
\begin{equation}
  \label{eq:neumann}
  [\mathcal{E},\Delta] \geq -\sum_{\alpha=1}^{N}\sigma_{\alpha}\tau_{\alpha}
\end{equation}
where the equality holds if and only if 
\begin{equation}
  \label{eq:cond-neumann}
(\mathbf{u}_{\alpha}^{T}\mathbf{x}_{\beta})
(\mathbf{v}_{\alpha}^{T}\mathbf{y}_{\beta}) = -\delta_{\alpha,\beta}
\end{equation}
for all $\alpha$ and $\beta$ with non-zero singular values
$\sigma_{\alpha}$ and $\tau_{\beta}$ ($\delta_{\alpha,\beta}$ is 
Kronecker's delta). 
We now regard this inequality as a lower bound for the conformational energy
for a given $E$-matrix. For a fixed set of the singular values $\sigma_{\alpha}$
($\alpha = 1, \cdots, N$), if and only if there exists such a conformation 
that satisfies the condition in Eq. (\ref{eq:cond-neumann}), then 
that conformation is the lowest possible energy conformation. 
Let $\lambda_{\alpha}$ and $\varepsilon_{\alpha}$ ($\alpha = 1,\cdots,N$) 
be the eigenvalues of the $C$-matrix and $E$-matrix, respectively, 
sorted in the decreasing order of their absolute values. Then
$\sigma_{\alpha} = |\lambda_{\alpha}|$ and $\tau_{\alpha} = |\varepsilon_{\alpha}|$
for $\alpha = 1, \cdots, N$, and $\mathbf{u}_{\alpha}$ and $\mathbf{x}_{\alpha}$
are the eigenvectors of the corresponding matrices.
Thus, in terms of eigenvalues and eigenvectors, 
the lower bound in Eq. (\ref{eq:neumann}) is equal to 
$\sum_{\alpha}\lambda_{\alpha}\varepsilon_{\alpha}$ with $\lambda_{\alpha}\varepsilon_{\alpha} \leq 0$ for $\alpha = 1, \cdots, N$. 
In addition to the condition Eq. (\ref{eq:cond-neumann}) for the lower bound of 
Eq. (\ref{eq:neumann}), if $\Delta$ and $\mathcal{E}$ are of the same rank,
then the numbers of positive, negative, and zero eigenvalues of 
$\Delta$ and $-\mathcal{E}$ are the same  
and $\mathbf{u}_{\alpha} = \pm\mathbf{x}_{\alpha}$. Thus, from Sylvester's 
law of inertia~\cite{MatrixAnalysis}, there exists a real non-singular 
matrix $S$ such that 
\begin{equation}
  \label{eq:sylvester}
  \mathcal{E} = -S\Delta S^{T},
\end{equation}
i.e., the $E$-matrix is $^*$congruent to the $C$-matrix.
If the conformation that satisfy the condition Eq.~(\ref{eq:sylvester})
is the native structure, the $E$-matrix is consistent in the 
sense that the contributions from all the eigencomponents are stabilizing 
the native structure ($\lambda_{\alpha}\varepsilon_{\alpha} \leq 0$).
Since the matrix $S$ is non-singular, we can ``predict'' the native structure
from the $E$ matrix as $\Delta = -S^{-1}\mathcal{E}S^{-T}$ (if we can construct
the appropriate matrix $S$). At this point, however, the native structure may 
not be the GMEC since other conformations with a different set of singular 
values may have lower energies.

In order to compare the energies of conformations with different sets of 
singular values, we use another inequality~\cite{TopicsInMatrixAnalysis}:
\begin{equation}
  \label{eq:cs-ineq2}
  -\sum_{\alpha=1}^{N}\sigma_{\alpha}\tau_{\alpha} 
\geq -\|\mathcal{E}\| \|\Delta\|
\end{equation}
where the lower bound is the same as that in Eq. (\ref{eq:cs-ineq}).
We note that, in terms of singular values, the matrix norms are expressed as 
$\|\Delta\| = (\sum_{\alpha}\sigma_{\alpha}^2)^{1/2}$ 
and $\|\mathcal{E}\| = (\sum_{\alpha}\tau_{\alpha}^2)^{1/2}$. 
Hence, it is clear that the equality in Eq. (\ref{eq:cs-ineq2}) holds if and 
only if, in addition to the condition in Eq. (\ref{eq:cond-neumann}), 
there exists a scalar constant $c$ such that 
$\tau_{\alpha} = c\sigma_{\alpha}$ for all 
$\alpha = 1, \cdots, N$. 
These conditions are equivalent to Eq. (\ref{eq:go-pot}).

\subsection{One-dimensional approximations}
To connect the present results with previous studies, we next introduce 
two approximations. 
First, we consider the case where the $E$-matrix is well 
approximated by its principal eigencomponent, that is, 
$\mathcal{E} \approx \varepsilon_{1}\mathbf{x}_1\mathbf{x}_{1}^T$.
This approximation is motivated by the eigenvalue analysis of the 
Miyazawa-Jernigan (MJ) contact potential~\cite{MiyazawaANDJernigan1985} 
performed by Li et al.~\cite{Li-Tang-Wingreen1997}, and has been employed by 
others~\cite{CaoETAL2004,CaoETAL2006,BastollaETAL2005}.
In this case, the lower bound Eq. (\ref{eq:neumann}) is achieved if and only if
$\mathbf{x}_1 = \pm\mathbf{u}_1$ and $\varepsilon_{1}\lambda_{1} < 0$.
This result was previously derived by Cao et al.~\cite{CaoETAL2004}
who subsequently showed that the vector $\mathbf{x}_1$ constructed by using 
the components of the principal eigenvector of the MJ contact potential
is indeed highly correlated with the principal eigenvector of the native 
contact matrices~\cite{CaoETAL2006}. Bastolla et al.~\cite{BastollaETAL2005} obtained a similar 
result, but they also showed that taking the average of such $\mathbf{x}_1$ 
over evolutionarily related proteins greatly improved the correlation.
Since the rank of the contact matrix is in general not 1,
Eq. (\ref{eq:sylvester}) does not hold and the equality in 
Eq. (\ref{eq:cs-ineq}) cannot be satisfied. 
Consequently, there are attractive interactions between non-native contacts 
even when $\mathbf{x}_1 = \mathbf{u}_1$ holds exactly.
Nevertheless, Porto et al.~\cite{PortoETAL2004} have demonstrated 
that the knowledge of $\mathbf{u}_1$ alone is practically sufficient for 
reconstructing the native contact matrix of small single-domain proteins. 
Therefore, construction of effective 
rank-1 $E$-matrices is of great interest~\cite{VulloETAL2006}.
Based on the Porto et al.'s result, it is tempting to postulate that the 
satisfaction of the lower bound by a rank-1 $E$-matrix is sufficient for 
the native conformation to be the unique GMEC.
At present, however, there is no clear connection between 
the present formulation (energy minimization) and 
the Porto et al.'s combinatorial algorithm.

Another approximation is a kind of mean-field approximations in which 
the matrix element $\mathcal{E}_{ij}$ is replaced by its average over column
$\langle \mathcal{E}_{i\bullet} \rangle \equiv \sum_{j=1}^{N}\mathcal{E}_{ij}/N$. 
Let us define $\mathbf{e} = (
\langle \mathcal{E}_{1\bullet} \rangle, \cdots, 
\langle \mathcal{E}_{N\bullet} \rangle)^T$ and 
$\mathbf{n} = (n_1, \cdots, n_N)^T$ where $n_i \equiv \sum_{j=1}^{N} \Delta_{ij}$
is the contact number of the $i$-th residue.
Then, we have the following approximation and the lower bound:
\begin{eqnarray}
  E(C,S) & \approx & \frac{1}{2}\mathbf{e}^T\mathbf{n}\\
\label{eq:cn-ineq}
& \geq & -\frac{1}{2}\| {\mathbf{e}}\| \|\mathbf{n}\|
\end{eqnarray}
where the equality in (\ref{eq:cn-ineq}) holds if and only if 
the column-averaged $E$-matrix is anti-parallel to the 
contact number vector, that is, ${\mathbf{e}} = 
\varepsilon\mathbf{n}$ for some $\varepsilon < 0$.
This lower bound condition is analogous to Eq. (\ref{eq:go-pot}), and 
can be regarded as another kind of the G\=o potential for one-dimensional
protein structure.
It has been suggested that contact number vector 
can significantly constrain the conformational space~\cite{KabakciogluETAL2002}.
Together with other one-dimensional structures, contact number vector is also 
used for recovering the native structures~\cite{KinjoANDNishikawa2005},
and can be accurately predicted~\cite{KinjoETAL2005,KinjoANDNishikawa2005c,Yuan2005,IshidaETAL2006,KinjoANDNishikawa2006}. 
It has been pointed out that the contact number vector is highly 
correlated with the principal eigenvector of the 
$C$-matrix~\cite{PortoETAL2004,KinjoANDNishikawa2005}, 
which suggests that this mean-field approximation is qualitatively similar to 
the principal eigenvector approximation introduced above.

\section{Discussion}
Using a more restricted, but conventional, form of the $E$-matrix where 
each element 
$\mathcal{E}_{ij}$ depends only on the types of $i$-th and $j$-th residues
(e.g., the MJ potential), 
Vendruscolo et al.~\cite{VendruscoloANDDomany1998,VendruscoloETAL2000} have 
shown that it is impossible for such $E$-matrices to stabilize all the native 
structures in a database. 
The conventional $E$-matrices such as those they studied do not take 
into account the sequence-dependence beyond a summation of the contributions 
from residue pairs.
In the present study, we assumed a more general form for the $E$-matrix, 
allowing each element $\mathcal{E}_{ij}$ to depend on the whole amino acid 
sequence. 
In practical situations of protein structure prediction, we want to optimize 
an energy function so that the native conformations of arbitrary proteins 
achieve the lower bound. 
Now let us impose this as a requisite for the $E$-matrix. Then, there should 
exist a function, namely $\cal{E}$, that maps each amino acid sequence 
to the corresponding optimal $E$-matrix, that is, the G\=o potential. 
Thus, the problem of structure prediction becomes a trivial matter. 
Currently, most efforts for developing energy functions seem to be focused on 
accurate estimation of a fixed set of parameters for a given functional 
form~\cite{HaoANDScheraga1999}.
The present analysis suggests that inferring the function $\cal{E}$ that 
can generate the G\=o-like $E$-matrices from amino acid sequences is 
essential if a contact potential is used. 
The lower bound inequality (Eq. \ref{eq:cs-ineq}) and its condition for 
the equality (Eq. \ref{eq:go-pot}) will serve as the guiding principle 
for inferring such a function. This approach to structure prediction is 
apparently similar to machine-learning approaches to contact matrix 
prediction~\cite{VulloETAL2006,ChengANDBaldi2007}. Although conventional 
machine-learning methods are not directly targeted at the optimization of 
the form of Eq. (\ref{eq:cs-ineq}), their prediction accuracy should be 
indicative of the possibility for identifying the function $\cal{E}$.

In the preceding paragraph, we have assumed the existence of the function
$\cal{E}$ to construct the optimal contact potential from a given amino acid 
sequence. What if, however, there is no such function? In fact, the limited 
success of current contact matrix prediction~\cite{CASP6_Contact} strongly 
suggests that this is more likely the case.
Such a case implies either that there are proteins for which the lower bound
energy cannot be achieved, or that the total energy cannot be sufficiently 
accurately approximated by Eq. (\ref{eq:ene}). 
The former case indicates that some proteins are inherently frustrated, but 
to a good approximation such proteins should be rather exceptional for 
natural proteins~\cite{Go1983,Takada1999}.
The latter case may indicate that multi-body contact 
interactions~\cite{MunsonANDSingh1997} and/or 
other energy components than contact energies are more important. 

In summary, we have shown that the requirement for the native structure to 
achieve the lower bound naturally leads to the G\=o potential and the 
requirement for such a conformation to be the unique GMEC leads to 
the native conformation being the most compact one among those containing all 
the native contacts.
These results suggest that protein structure prediction should be possible 
simply by constructing the optimal energy matrices or that the contact 
potential alone is not suitable for the problem. Although not yet definitive, 
the current state of contact prediction~\cite{CASP6_Contact} as well as recent 
studies on local interactions~\cite{ChikenjiETAL2006,FlemingETAL2006} suggest 
that the latter may be the case. Nevertheless, the present results may be
useful for evaluating the optimality of potential functions in either case.

\begin{thebibliography}{10}
\expandafter\ifx\csname url\endcsname\relax
  \def\url#1{\texttt{#1}}\fi
\expandafter\ifx\csname urlprefix\endcsname\relax\def\urlprefix{URL }\fi

\bibitem{Anfinsen1973}
C.~B. Anfinsen, Principles that govern the folding of protein chains, Science
  181 (1973) 223--230.

\bibitem{HaoANDScheraga1999}
M.-H. Hao, H.~A. Scheraga, Designing potential energy functions for protein
  folding, Curr. Opin. Struct. Biol. 9 (1999) 184--188.

\bibitem{MitsutakeETAL2001}
A.~Mitsutake, Y.~Sugita, Y.~Okamoto, Generalized-ensemble algorithms for
  molecular simulations of biopolymers, Biopolymers 60 (2001) 96--123.

\bibitem{MiyazawaANDJernigan1985}
S.~Miyazawa, R.~L. Jernigan, Estimation of effective interresidue contact
  energies from protein crystal structures: quasi-chemical approximation,
  Macromolecules 18 (1985) 534--552.

\bibitem{MiyazawaANDJernigan1996}
S.~Miyazawa, R.~L. Jernigan, Residue-residue potentials with a favorable
  contact pair term and an unfavorable high packing density term for simulation
  and threading, J. Mol. Biol. 256 (1996) 623--644.

\bibitem{VendruscoloETAL1997}
M.~Vendruscolo, E.~Kussell, E.~Domany, Recovery of protein structure from
  contact maps, Fold. Des. 2 (1997) 295--306.

\bibitem{TaketomiETAL1975}
H.~Taketomi, Y.~Ueda, N.~G\=o, Studies on protein folding, unfolding and
  fluctuations by computer simulation. {I}. the effect of specific amino acid
  sequence represented by specific inter-unit interactions, Int. J. Pept.
  Protein Res. 7 (1975) 445--459.

\bibitem{MatrixAnalysis}
R.~A. Horn, C.~R. Johnson, Matrix analysis, Cambridge University Press,
  Cambridge, U. K., 1985.

\bibitem{TopicsInMatrixAnalysis}
R.~A. Horn, C.~R. Johnson, Topics in matrix analysis, Cambridge University
  Press, Cambridge, U. K., 1991.

\bibitem{Go1983}
N.~G{\={o}}, Theoretical studies of protein folding, Annu. Rev. Biophys.
  Bioeng. 12 (1983) 183--210.

\bibitem{Takada1999}
S.~Takada, {\em G{\=o}}-ing for the prediction of protein folding mechanisms,
  Proc. Natl. Acad. Sci. U.S.A. 96 (1999) 11698--11700.

\bibitem{Li-Tang-Wingreen1997}
H.~Li, C.~Tang, N.~S. Wingreen, Nature of driving force for protein folding: A
  result from analyzing the statistical potential, Phys. Rev. Lett. 79 (1997)
  765--768.

\bibitem{CaoETAL2004}
H.~B. Cao, Y.~Ihm, C.~Z. Wang, M.~Su, D.~Dobbs, K.~M. Ho, Three-dimensional
  threading approach to protein structure recognition, Polymers 45 (2004)
  687--697.

\bibitem{CaoETAL2006}
H.~B. Cao, C.~Z. Wang, D.~Dobbs, Y.~Ihm, K.~M. Ho, Codability criterion for
  picking proteinlike structures from random three-dimensional configurations,
  Phys. Rev. E 74 (2006) 031921.

\bibitem{BastollaETAL2005}
U.~Bastolla, M.~Porto, H.~E. Roman, M.~Vendruscolo, Principal eigenvector of
  contact matrices and hydrophobicity profiles in proteins, Proteins 58 (2005)
  22--30.

\bibitem{PortoETAL2004}
M.~Porto, U.~Bastolla, H.~E. Roman, M.~Vendruscolo, Reconstruction of protein
  structures from a vectorial representation, Phys. Rev. Lett. 92 (2004)
  218101.

\bibitem{VulloETAL2006}
A.~Vullo, I.~Walsh, G.~Pollastri, A two-stage approach for improved prediction
  of residue contact map, BMC Bioinformatics 7 (2006) 180.

\bibitem{KabakciogluETAL2002}
A.~Kabak\c{c}io\v{g}lu, I.~Kanter, M.~Vendruscolo, E.~Domany, Statistical
  properties of contact vectors, Phys. Rev. E 65 (2002) 041904.

\bibitem{KinjoANDNishikawa2005}
A.~R. Kinjo, K.~Nishikawa, Recoverable one-dimensional encoding of
  three-dimensional protein structures, Bioinformatics 21 (2005) 2167--2170,
  doi:10.1093/bioinformatics/bti330.

\bibitem{KinjoETAL2005}
A.~R. Kinjo, K.~Horimoto, K.~Nishikawa, Predicting absolute contact numbers of
  native protein structure from amino acid sequence, Proteins 58 (2005)
  158--165, doi:10.1002/prot.20300.

\bibitem{KinjoANDNishikawa2005c}
A.~R. Kinjo, K.~Nishikawa, Predicting secondary structures, contact numbers,
  and residue-wise contact orders of native protein structure from amino acid
  sequence using critical random networks, BIOPHYSICS 1 (2005) 67--74,
  doi:10.2142/biophysics.1.67.

\bibitem{Yuan2005}
Z.~Yuan, Better prediction of protein contact number using a support vector
  regression analysis of amino acid sequence, BMC Bioinformatics 6 (2005) 248.

\bibitem{IshidaETAL2006}
T.~Ishida, S.~Nakamura, K.~Shimizu, Potential for assessing quality of protein
  structure based on contact number prediction, Proteins 64 (2006) 940--947.

\bibitem{KinjoANDNishikawa2006}
A.~R. Kinjo, K.~Nishikawa, {CRNPRED}: Highly accurate prediction of
  one-dimensional protein structures by large-scale critical random networks,
  BMC Bioinformatics 7 (2006) 401.

\bibitem{VendruscoloANDDomany1998}
M.~Vendruscolo, E.~Domany, Pairwise contact potentials are unsuitable for
  protein folding, J. Chem. Phys. 109 (1998) 11101--11108.

\bibitem{VendruscoloETAL2000}
M.~Vendruscolo, R.~Najmanovich, E.~Domany, Can a pairwise contact potential
  stabilize native protein folds against decoys obtained by threading?,
  Proteins 38 (2000) 134--148.

\bibitem{ChengANDBaldi2007}
J.~Cheng, P.~Baldi, Improved residue contact prediction using support vector
  machines and a large feature set, BMC Bioinformatics 8 (2007) 113.

\bibitem{CASP6_Contact}
O.~Gra\~na, D.~Baker, R.~M. MacCallum, J.~Meiler, M.~Punta, B.~Rost, M.~L.
  Tress, A.~Valencia, {CASP6} assessment of contact prediction, Proteins Suppl.
  7 (2005) 214--224.

\bibitem{MunsonANDSingh1997}
P.~J. Munson, R.~K. Singh, Statistical significance of hierarchical multi-body
  potentials based on delaunay tessellation and their application in
  sequence-structure alignment, Protein Sci. 6 (1997) 1467--1481.

\bibitem{ChikenjiETAL2006}
G.~Chikenji, Y.~Fujitsuka, S.~Takada, Shaping up the protein folding funnel by
  local interaction: lesson from a structure prediction study, Proc. Natl.
  Acad. Sci. U.S.A. 103 (2006) 3141--3146.

\bibitem{FlemingETAL2006}
P.~J. Fleming, H.~Gong, G.~D. Rose, Secondary structure determines protein
  topology, Protein Sci. 15 (2006) 1829--1834.

\end{thebibliography}
\bibliographystyle{elsart-num}

\end{document}